\documentclass[11pt,letterpaper]{article}
\usepackage[utf8]{inputenc}
\usepackage{multirow} 
\usepackage{booktabs} 
\usepackage[english]{babel}
\usepackage{amsmath}
\usepackage{amsfonts}
\usepackage{amssymb}
\usepackage{graphicx}
\usepackage[left=3cm,right=3cm,top=2cm,bottom=3cm]{geometry}
\author{Jorge Pinochet}
\title{\textbf{Five misconceptions about black holes}}
\begin{document}

\author{Jorge Pinochet\\ \\
 \small{\textit{Facultad de Educación}}\\
 \small{\textit{Universidad Alberto Hurtado, Erasmo Escala 1835, Santiago, Chile.} japinochet@gmail.com}\\}

\date{}
\maketitle

\begin{center}\rule{0.9\textwidth}{0.1mm} \end{center}
\begin{abstract}
\noindent Given the great interest that black holes arouse among non-specialists, it is important to analyse misconceptions related to them. According to the author, the most common misconceptions are that: (1) black holes are formed from stellar collapse; (2) they are very massive; (3) they are very dense; (4) their gravity absorbs everything; and (5) they are black. The objective of this work is to analyse and correct these misconceptions. This article may be useful as pedagogical material in high school physics courses or in introductory courses in undergraduate physics. \\ \\

\noindent \textbf{Keywords}: Black holes, misconceptions, high school students, science-engineering undergraduate students. 

\begin{center}\rule{0.9\textwidth}{0.1mm} \end{center}
\end{abstract}

\maketitle

\section{Introduction}

Popular science books and physics education papers about black holes usually focus on analysing the properties of these objects, but pay little or no attention to misconceptions related to them. Given the great interest that black holes arouse among non-specialists, which has been increased by the first image of a black hole recently disseminated throughout the world by the mass media, it is nowadays especially important to analyse these misconceptions. This is especially relevant in the case of physics teachers, both at school and university level, who among other challenges have the responsibility to correct the scientific misconceptions of their students.\\

According to the author, the most common misconceptions are that: (1) black holes are formed from stellar collapse; (2) they are very massive; (3) they are very dense; (4) their gravity absorbs everything; and (5) they are black. This last misconception is perhaps the most surprising, since it contradicts the very definition of a black hole. The objective of this work is to analyse and correct these misconceptions. Given the educational purpose of this article, we will adopt an intuitive approach based on introductory physics. In addition, we will focus on the simplest black hole, known as a static black hole\footnote{In theory, there are three classic externally observable parameters that define a black hole: mass, electrical charge and angular momentum. The simplest black hole is then one that lacks rotation and is electrically neutral. This object is the static black hole, also called Schwarzschild black hole, and its mathematical description only depends on its mass.}  [1]. These simplifications will not affect the validity of the conclusions, and will allow us to sidestep the complex notions of Einstein's theory of general relativity, which is the appropriate framework for a technical analysis. These five misconceptions are analysed separately in Sections 2 to 6, in the order indicated above. Section 2 also serves as a brief introduction to the concept of a black hole. The article ends with a brief summary and discussion.

\section{Black holes are formed from stellar collapse}

The concept of a black hole was born within the framework of general relativity, which is the theory of gravity proposed by Einstein in 1916 to extend and perfect Newton's law of universal gravitation [2, 3]. A black hole is a region of space-time limited by a closed surface called the \textit{horizon}, where there is such a high concentration of mass-energy that nothing can escape its gravity, not even light. In the framework of Newtonian gravitation, the horizon can be intuitively visualised as a spherical surface of radius\footnote{In order to avoid introducing another misconception, it is important to keep in mind that in general relativity, $R_{S}$ is a coordinate, and therefore does not represent a physical distance.} [4, 5]:

\begin{equation} 
R_{S} = \frac{2GM_{BH}}{c^{2}} = 1.48 \times 10^{-27} m \left( \frac{M_{BH}}{kg} \right). 
\end{equation}

For historical reasons, this is called the \textit{Schwarzschild radius}, where $G = 6.67 \times 10^{-11} N\cdot m^{2} \cdot kg^{-2}$ is the gravitational constant, $c = 3 \times 10^{8} m \cdot s^{-1}$ is the speed of light in vacuum, and $M_{BH}$ is the mass of the black hole, that is, the mass enclosed within the horizon. Eq. (1) tells us that for each value of $M_{BH}$, there is a value of $R_{S}$, meaning that in principle there is no restriction on the mass that a black hole can have. Hence, while it is true that a black hole can be formed from a star, it can also theoretically be formed from a planet, an asteroid or a grain of sand, provided that the raw material for the formation process is ultimately confined within the corresponding horizon [1, 6]. However, only at the origins of the universe (when the average density of mass-energy was very high) were suitable conditions created for the formation of black holes from comparatively small concentrations of matter, such as those equivalent to an asteroid or a grain of sand.\\

Under the conditions that currently prevail in the universe, the formation of a black hole requires a mass that is at least as large as that of a star. However, a star cannot spontaneously form a black hole, since its high internal temperatures prevent this. In essence, what happens is that the compressive force of gravity is counteracted by the expansive thermal pressure generated by the thermonuclear reactions occurring in the centre of the star. Nevertheless, if the star is massive enough, the gravity exerted on itself is so powerful that once the nuclear fuel has been exhausted and the thermal pressure drops, a gravitational collapse ensues that no force in nature can stop, and as a result, a \textit{stellar-mass black hole} is formed.\\

It is estimated that the minimum mass needed by an object that is not undergo nuclear reactions for gravity to impose itself on all other forces is $\sim3M_{\odot}$, where $M_{\odot} = 1.99 \times 10^{30} kg$ is the solar mass. This figure is known as the \textit{Tolman-Oppenheimer-Volkoff limit} (TOV) in honour of the physicists who calculated it for the first time [7, 8]. Then, an object that is not undergo nuclear reactions whose mass exceeds $\sim3M_{\odot}$ must collapse gravitationally to form a black hole \footnote{To be precise, the TOV limit is an upper bound to the mass of stars composed mainly of degenerate neutrons (neutron stars). This limit is subject to uncertainty, because the equation of state for hadronic matter subjected to high densities is still not well known.}.\\

Table 1 shows the Schwarzschild radius (calculated from Eq. (1)) and other quantities for different objects, expressed in powers of 10. The same procedure is also used in the tables in the following sections.

\begin{table}[htbp]
\begin{center}
\caption{Schwarzschild radius of some objects}
\begin{tabular}{l l l l} 
\toprule
Object & Mass ($kg$) & Typical radius ($m$) & $R_{S}$ ($m$)\\
\midrule
Sun & $10^{30}$ & $10^{8}$ & $10^{3}$ \\ 
\midrule
Earth & $10^{24}$ & $10^{6}$ & $10^{-3}$ \\ 
\midrule
Asteroid (average) & $10^{12}$ & $10^{3}$ & $10^{-15}$ \\
\midrule
Grain of sand & $10^{-6}$ & $10^{-3}$ & $10^{-33}$ \\
\bottomrule
\end{tabular}
\label{Schwarzschild radius of some objects}
\end{center}
\end{table}

\section{Black holes are very massive} 

Table 1 reveals that black holes can have a wide range of masses, and are therefore not necessarily massive. As pointed out in Section 2, what characterises a black hole is that its mass, whether small or large, is confined within a horizon. Technically, it can be said that a black hole is a \textit{compact object}; this is a property that is different from the density, which we will analyse in the next section. The compactness does not depend on the mass taken in isolation, but on the quotient of the mass and the radius. Specifically, the compactness of a spherical object of mass $M$ can be defined as the quotient of its Schwarzschild radius $R_{S}$ and its real radius $R$:

\begin{equation} 
\frac{R_{S}}{R} = \frac{2GM}{c^{2} R}.
\end{equation}

The closer to unity this quotient is, the more compact the object is. If we take $R_{S}$ as a measure of the size of a black hole, then for these objects it is true that $2GM/c^{2} R =1$. This means that black holes are the most compact objects in the universe [9], and it is intuitively evident that all other objects must satisfy the relationship:

\begin{equation} 
\frac{2GM}{c^{2} R} < 1.
\end{equation}

This result is known as the \textit{Buchdahl inequality}\footnote{More precisely, the Buchdahl inequality states that, under certain technical conditions, a sphere of static fluid of mass $M$ and radius $R$ satisfies the relation $2GM/c^{2}R < 8/9$.}, and is rigorously derived from general relativity [10]. Table 2 shows the compactness (expressed as the $R_{S}/R$ ratio) of different objects.

\begin{table}[htbp]
\begin{center}
\caption{Compactness of several objects (for a black hole, $R_{S}/R = 1$)}
\begin{tabular}{l l l l l} 
\toprule
Object & Mass ($kg$) & Typical radius ($m$) & $R_{S}$ ($m$) & $R_{S}/R$\\
\midrule
Neutron star & $10^{30}$ & $10^{4}$ & $10^{3}$ & $10^{-1}$ \\ 
\midrule
White dwarf & $10^{30}$ & $10^{6}$ & $10^{3}$ & $10^{-3}$ \\ 
\midrule
Sun & $10^{30}$ & $10^{8}$ & $10^{3}$ & $10^{-5}$ \\
\midrule
Earth & $10^{24}$ & $10^{6}$ & $10^{-3}$ & $10^{-9}$ \\
\bottomrule
\end{tabular}
\label{Compactness of several objects}
\end{center}
\end{table}

\section{Black holes are very dense}

Contrary to what common sense suggests, the density of a black hole is not necessarily high, and under certain conditions it can be quite low. Since the only size that we can assign to a black hole is determined by its Schwarzschild radius, we can define its density $\rho_{BH}$ as the ratio between its mass and the volume of a sphere of radius $R_{S}$:

\begin{equation} 
\rho_{BH} = \frac{M_{BH}}{4\pi R^{3}_{S}/3}.
\end{equation}

We can interpret this quantity as an average density, since when an object has contracted to beyond its Schwarzschild radius, this contraction continues without anything being able to stop it, meaning that the density in the centre of the horizon increases without limit [1]. Introducing Eq. (1) into (4):

\begin{equation} 
\rho_{BH} = \frac{3c^{6}}{32\pi G^{3}M_{BH}^{2}} = 7.3 \times 10^{79} kg\cdot m^{-3} \left( \frac{kg}{M_{BH}} \right)^{2}.
\end{equation}

\begin{table}[htbp]
\begin{center}
\caption{Average densities of different types of black hole}
\begin{tabular}{l l l l} 
\toprule
Object & $M_{BH}$ ($kg$) & $M_{BH}$ ($M_{\odot}$) & $\rho_{BH}$ ($kg\cdot m^{-3}$)\\
\midrule
Supermassive black hole & $10^{36} - 10^{40}$ & $10^{6} - 10^{10}$ & $10^{7} - 10^{-1}$ \\ 
\midrule
Intermediate-mass black hole & $10^{33} - 10^{35}$ & $10^{3} - 10^{5}$ & $10^{13} - 10^{9}$ \\ 
\midrule
Stellar-mass black hole & $10^{30} - 10^{32}$ & 1 - $10^{12}$ & $10^{19} - 10^{15}$ \\
\midrule
Micro-black hole & $10^{12} - 10^{-8}$ & $10^{-18} -10^{-38}$ & $10^{55} - 10^{95}$ \\
\bottomrule
\end{tabular}
\label{Average densities of different types of black hole}
\end{center}
\end{table}

Table 3 shows values of $\rho_{BH}$ for different types of black hole, including the mass ranges in $kg$ and units of solar mass, $M_{\odot}$ for each. In addition to the stellar-mass black hole and the supermassive black hole, the only types for which there is solid observational evidence, two hypothetical objects are also shown: the \textit{intermediate-mass black hole} and the \textit{micro-black hole}\footnote{These two kinds of black hole are both hypothetical, but their astronomical status is different. Although there is no evidence for micro-black holes, indications of the existence of intermediate-mass black holes have emerged in recent years. The most recent come from observations made in Chile with the ALMA telescope, where evidence was found of a very compact object of mass $10^{4}M_{\odot}$ in the Sirius constellation [11].}. In theory, the latter could have been formed shortly after the big bang, when the mass-energy density was very high.\\

As a basis for comparison, remember that the density of fresh water on the earth's surface is $10^{3} kg\cdot m^{-3}$. As indicated by Eq. (5) and Table 3, the average density of the largest supermassive black holes that have been observed may be less than that of the fresh water on Earth.

\section{The gravity of a black hole absorbs everything}

The gravity outside a spherical star of mass $M$ is identical to that generated by a black hole of the same mass . However, we know that stars do not absorb everything around them, and this is also therefore true of black holes. In fact, planets have stable orbits around stars for billions of years; this means that if the Sun were replaced by a black hole of mass $M_{\odot}$, we would not notice anything anomalous except for the obvious lack of light. In other words, the Earth would continue to describe a stable orbit around our star. In addition, at the distance of the Earth from the sun, the effects of general relativity are practically indistinguishable from those predicted by the law of Newtonian gravitation.\\

However, this scenario changes dramatically when a body is too close to a black hole. In general relativity, the \textit{innermost stable circular orbit} (ISCO) is the smallest orbit in which a test particle can stably orbit a black hole [1, 9] (see Table 4). In the case of a Schwarzschild hole, the ISCO radius is calculated as [1]:

\begin{equation} 
R_{ISCO} = 3R_{S} = \frac{6GM_{BH}}{c^{2}} = 4.4\times 10^{-27} m \left( \frac{M_{BH}}{kg} \right).
\end{equation}

This radius is extremely small in relation to the typical size of a star. For example, if we take $M_{BH} = M_{\odot}$ we get $R_{ISCO} \sim 10^{3}m$, which represents only one hundred thousandth of the average solar radius, $R_{\odot} \sim 10^{8}m$.\\

Below $R_{ISCO}$, a particle cannot maintain a stable orbit, and is doomed to fall into the black hole and to be absorbed by it. This phenomenon has no equivalent in Newtonian physics, where there is always a stable orbit. In Einstein’s physics, this instability can be explained by recalling that a particle whose orbit is close to the horizon will have a relativistic velocity (close to $c$), which leads to a great increase in kinetic energy; based on the equivalence between mass and energy, this implies an increase in the mass of the particle that makes the gravitational attraction more intense, causing the particle to be finally absorbed.

\begin{table}[htbp]
\begin{center}
\caption{Innermost stable circular orbits for different types of black hole}
\begin{tabular}{l l} 
\toprule
Object & $R_{ISCO}$ ($m$) \\
\midrule
Supermassive black hole & $10^{9} - 10^{13}$ \\ 
\midrule
Intermediate-mass black hole & $10^{6} - 10^{8}$ \\ 
\midrule
Stellar-mass black hole & $10^{3} - 10^{5}$ \\
\midrule
Micro-black hole & $10^{-15} - 10^{-35}$ \\
\bottomrule
\end{tabular}
\label{Innermost stable circular orbits for different types of black hole}
\end{center}
\end{table}

\section{Black holes are black}

In analysing the misconceptions in the previous sections, we have used only ideas from general relativity and Newtonian gravitation. However, the analysis of the last erroneous concept requires the incorporation of quantum theory. The first physicist to present a detailed description of black holes that combined general relativity with quantum theory was Stephen Hawking, in a famous work published in 1974 and expanded in 1975\footnote{Strictly speaking, Hawking used \textit{quantum field theory}, which is a mathematical scheme that describes fundamental particles and their interactions (not including gravity), and that combines special relativity with ordinary quantum mechanics.} [12,13].\\

Hawking’s revolutionary discovery was that an isolated black hole has a temperature and emits thermal radiation from the horizon in all directions [14, 15]. In Hawking's words, "Black holes ain't so black" [16]. The fact that the black hole is isolated is important, because it means that the emission of radiation does not depend on mechanisms related to the absorption of material located outside the horizon, as happens with accretion.\\

Hawking showed that the thermal radiation emitted from the horizon of a black hole of mass $M_{BH}$, known as \textit{Hawking radiation}, has a blackbody spectrum with an absolute temperature given by [17]:

\begin{equation} 
T_{BH} = \frac{\hbar c^{3}}{8\pi kGM_{BH}} = 1.23\times 10^{23} K \left( \frac{kg}{M_{BH}} \right). 
\end{equation}

This is the so-called \textit{Hawking temperature}\footnote{The presence of the characteristic constants of general relativity ($c$ and $G$), quantum mechanics ($\hbar$) and thermodynamics ($k$) shows that this equation is obtained by combining these three theories.}, where $\hbar = h/2\pi = 1.05\times 10^{-34} J\cdot s$ is the reduced Planck constant, and $k = 1.38 \times 10^{-23} J\cdot K^{-1}$ is the Boltzmann constant. According to Hawking's calculations, $T_{H}$ is the temperature that would be registered by an observer located at a great distance from a black hole (ideally infinite). The inverse proportionality between $T_{H}$ and $M_{BH}$ in Eq. (7) reveals that Hawking radiation is only significant for small and light black holes such as micro-black holes. However, as we know, these objects have not been detected, and although there is no solid empirical evidence for the existence of Hawking radiation, for reasons of consistency with other physical theories that have been widely confirmed, specialists agree that Hawking radiation exists. 

\begin{table}[htbp]
\begin{center}
\caption{Hawking temperature for different types of black hole}
\begin{tabular}{l l l l} 
\toprule
Object & $T_{H}$ ($K$) & $\lambda_{max}$ ($m$) & Type of radiation\\
\midrule
Supermassive black hole & $10^{-13} - 10^{-17}$ & $10^{10} - 10^{14}$ & Radio waves \\ 
\midrule
Intermediate-mass black hole & $10^{-10} - 10^{-12}$ & $10^{7} - 10^{9}$ & Radio waves \\ 
\midrule
Stellar-mass black hole & $10^{-7} - 10^{-9}$ & $10^{4} - 10^{6}$ & Radio waves \\
\midrule
Micro-black hole & $10^{11} - 10^{31}$ & $10^{-14} -10^{-34}$ & Gamma rays \\
\bottomrule
\end{tabular}
\label{Hawking temperature for different types of black hole}
\end{center}
\end{table}

Table 5 shows different $T_{H}$ values (calculated from Eq. (7)), the corresponding radiation types and their wavelengths. It can be seen that in the case of micro-black holes, $T_{H}$ is very high and the horizon emits mainly gamma rays. However, in the case of supermassive black holes and stellar-mass black holes (the only ones for which there is observational evidence), $T_{H}$ is very close to absolute zero and Hawking radiation is undetectable.

\section{Final comments}

Black holes are no longer an issue of interest exclusively to specialists, and increasing numbers of people are becoming interested in discovering their secrets, which highlights the importance of correcting misconceptions related to them. From this perspective, I hope that the ideas addressed in this work contribute to a better understanding of black holes.\\

It is important to keep in mind that black hole physics is a very active research field that is full of questions and is in a permanent state of transformation. It should not therefore be surprising that in the near future, some of the ideas that are currently accepted about black holes will be rejected by theoretical or observational evidence, and end up becoming misconceptions, as for example with the belief, widely accepted until 1974, that black holes are black. However, these new ideas will be built on the current ones, and in order to understand and appreciate the advances of tomorrow we must be aware of the misconceptions of the present.

\section*{Acknowledgments}
I would like to thank to Daniela Balieiro and Michael Van Sint Jan for their valuable comments in the writing of this paper. 

\section*{References}

[1] V.P. Frolov, A. Zelnikov, Introduction to Black Hole Physics, Oxford University Press, Oxford, 2011.

\vspace{2mm}

[2] A. Einstein, Die Grundlage der allgemeinen Relativitätstheorie, Annalen der Physik, 354 (1916) 769-822.

\vspace{2mm}
[3] A. Einstein, The Collected Papers of Albert Einstein, Princeton University Press, Princeton, 1997.

\vspace{2mm}
[4] K.L. Lang, Essential Astrophysics, Springer, Berlin, 2013.

\vspace{2mm}
[5] D. Maoz, Astrophysics in a Nutshell, 2 ed., Princeton University Press, Princeton, 2016.

\vspace{2mm}
[6] V.P. Frolov, I.D. Novikov, Black Hole Physics: Basic Concepts and New Developments, Springer Science, Denver, 1998.

\vspace{2mm}
[7] J.R. Oppenheimer, G.M. Volkoff, On Massive Neutron Cores, Physical Review, 55 (1939) 374-381.

\vspace{2mm}
[8] R.C. Tolman, Static Solutions of Einstein's Field Equations for Spheres of Fluid, Physical Review, 55 (1939) 364-373.

\vspace{2mm}
[9] B. Schutz, A First Course in General Relativity, 2 ed., Cambridge University Press, Cambridge, 2009.

\vspace{2mm}
[10] G.A. Buchdahl, General Relativistic Fluid Spheres, Phys. Rev., 116 (1959) 1027-1034.

\vspace{2mm}
[11] S. Takekawa, et al. Indication of Another Intermediate-mass Black Hole in the Galactic Center, The Astrophysical Journal Letters, 871 L1 (2019).

\vspace{2mm}
[12] S.W. Hawking, Black Hole explosions?, Nature, 248 (1974) 30-31.

\vspace{2mm}
[13] S.W. Hawking, Particle creation by black holes, Communications in Mathematical Physics, 43 (1975) 199-220.

\vspace{2mm}
[14] J. Pinochet, Stephen Hawking y los Agujeros Negros Cuánticos, Rev. Mex. Fís. E, XX (2019).

\vspace{2mm}
[15] J. Pinochet, “Black holes ain't so black”: An introduction to the great discoveries of Stephen Hawking, Phys. Educ., XX (2019).

\vspace{2mm}
[16] S.W. Hawking, A brief history of time, Bantam Books, New York, 1998.

\vspace{2mm}
[17] J. Pinochet, The Hawking temperature, the uncertainty principle and quantum black holes, Phys. Educ., 53 065004 (2018).

\end{document}